Author: Mladen Georgiev (Institute of Solid State Physics, Bulgarian Academy of Sciences, 1784 Sofia, Bulgaria)

A current series of papers on barrier-controlled and trapping processes in solids and/or at solid surfaces have laid down the emphasis onto describing the statistical event by means of the barrier currents method due to Bardeen and Christov. The present study centred on photo-desorption and photo-electrification is the fourth of a series which also included rapid statistical approaches to nucleation, diffusion, and drift currents in state of the art materials. We presently extend our arguments to spherical square wells and to spherical oscillatory wells to deal with the photo-processes.

## 1. Foreword

Phonon (*alias* boson)-coupled processes are known to play a leading role in solid state (*alias* nuclear) physics. Some of these processes involve transitions across a barrier, others are associated with the coupling to a potential well. The former comprise diffusion, drift currents, nucleation, etc., the latter include various harmonic oscillators or trapping wells. However. other processes appear as pairs of the former two: motion across the central barrier of double-well potentials in small-polaron and chemical reactions. Other pairs of barriers and wells will become apparent shortly.

Laser sputtering (*alias* kind of photo-desorption) is amongst the variety of phenomena not only with essential applications to manufacturing clean materials for microelectronics but also with implications for their physical context. This turns laser sputtering into a desired playground for checking various physical models. One of them concerns the model electronic potential involved in sputtering (Itoh&Nakayama (I&N) potential) which is described as a flat bottom well at short distances followed by a screened repulsive Coulomb tail at longer ones [1]. There is a semi-barrier in between which controls the process of ejecting a bi-hole from the solid surface, which is the prerequisite for sputtering an atom.

The laser sputtering of atoms occurs when two photo-holes reside on a surface bond to destroy it. The I&N potential is therefore the electrostatic potential experienced by the second hole after the first hole has resided on a bond [1]. The Debye screening is very essential for producing a finite (concentration-dependent) semi-barrier thereby changing the effect on the second hole from increased repelling, as the partner hole moves towards the bond atom, to barrier controlled attraction when it moves away, as shown in Figure 1.

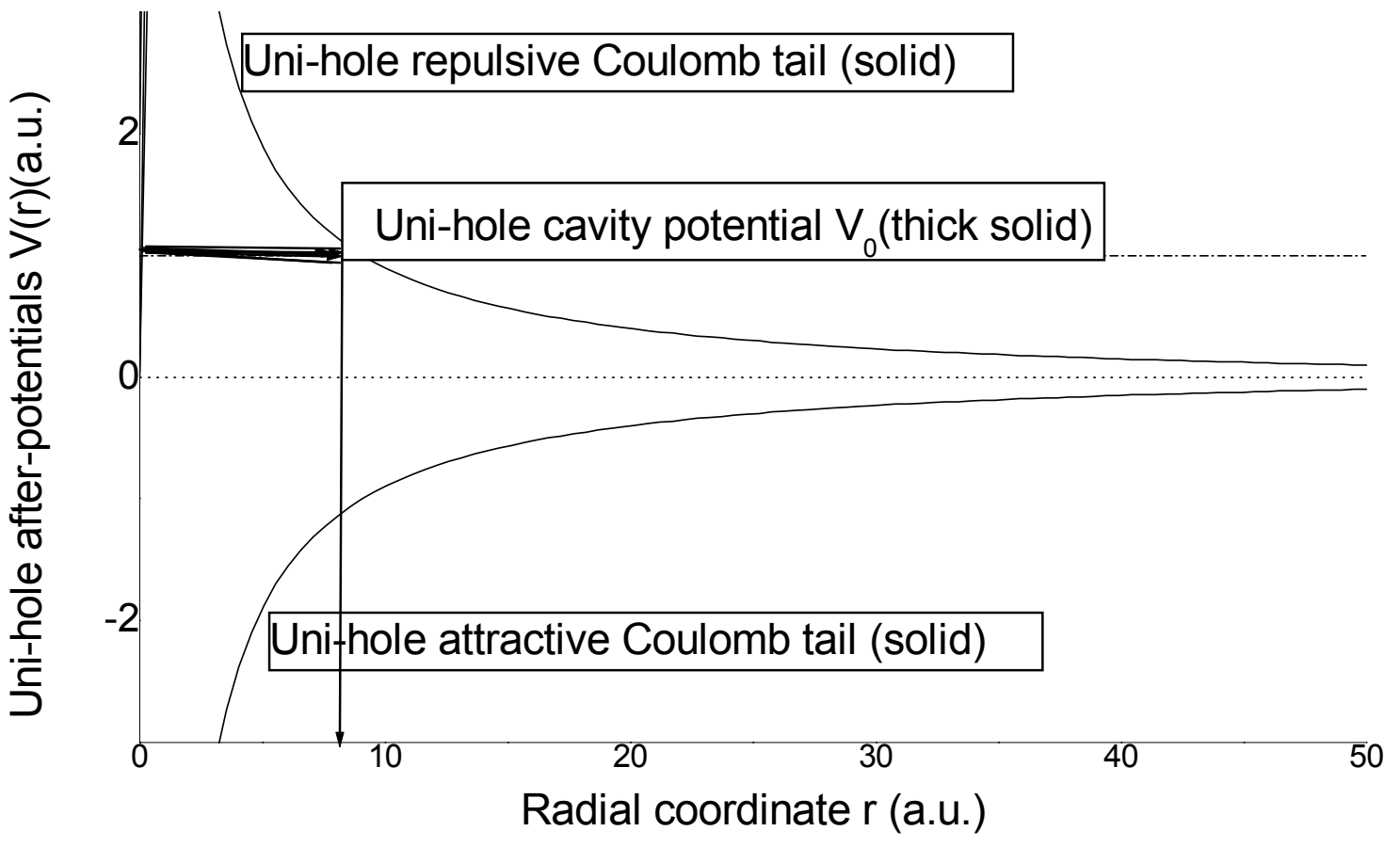

Figure 1:

Spherical potentials of frequent use for model calculations: (a)- Itoh-Nakayama's potential assumed to control the laser sputtering in $A_{III}B_V$ and (b)- the semi-continuum F center potential. In both of these the spherical cavity at short distances from the bond atom is superseded by a Coulomb tail at longer distances. The tail is screened repulsive in (a) and attractive in (b). In (a), Debye screening is essential for producing a concentration-dependent semi-barrier to control the second hole capture at the bond. See text for details.

If the sputtering particles are electrically charged (ions), then the sputtering process leads to charging the surface of the solid and to creating an ionic atmosphere outside it. In other words, sputtering of ions would lead to photo-electrification. In so far as only small amounts of ions are involved, the photo-voltages produced should fall into the microvolt range. Thus we have an example of a micro-scale voltage obtained from a normal-size body, alternative to nano-processes which involve signals from a nano-scale entity.

For an atom to be ejected from the surface by the above-described mechanism, the barrier process is likely to be phonon coupled [2]. Otherwise, the sputtering rate would not have been temperature-dependent as it appears to be. Extensions to a multimode picture being straightforward, we will confine our discussion to coupling to a single mode [3], as suggested by Figure 2.

The semi-barrier assumed to control I&N's process forms between the edge of the spherical well at $r = r_0$ and the ingredient of the screened potential about the coordinate axes. The active barrier is confined by the energy level traversed along by the approaching hole above the energy reference, as in Figure 3. The barrier width is thus limited by one of the energy levels of the Boltzmann tail statistics, depending on the temperature, excitation level, etc. In an alternative model proposed by Sumi the barrier width is dependent on the Fermi-Dirac statistics which allows for only one hole at a time to traverse along a definite band level [14]. Thus the barrier peak is always at r ~ $r_0$ but its width is statistics dependent.

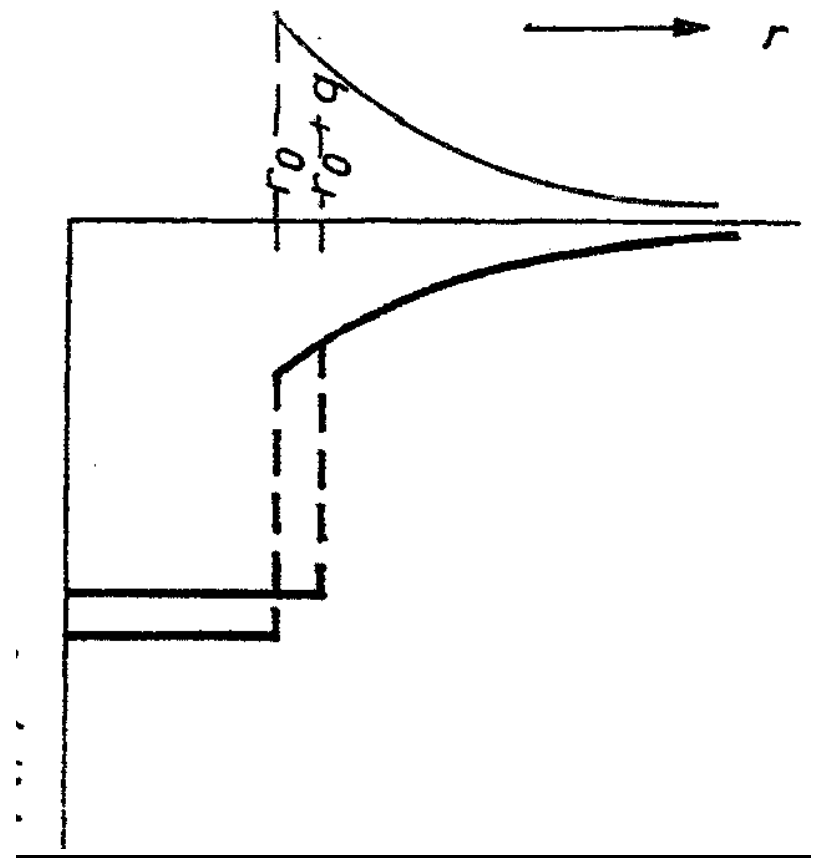

(a)

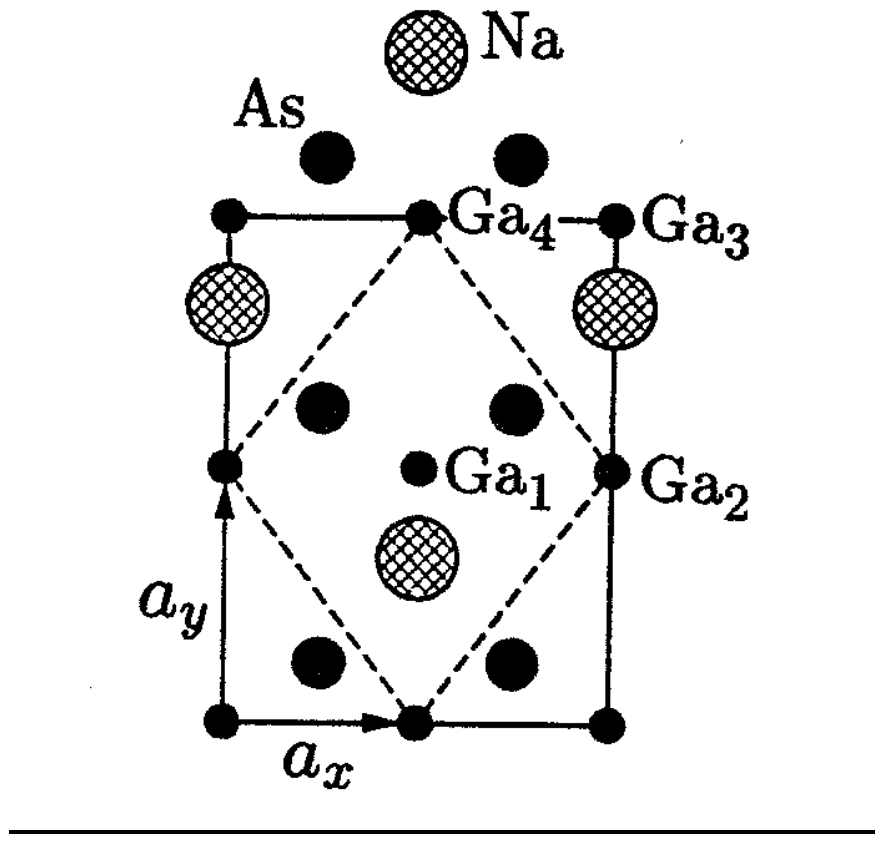

(b)

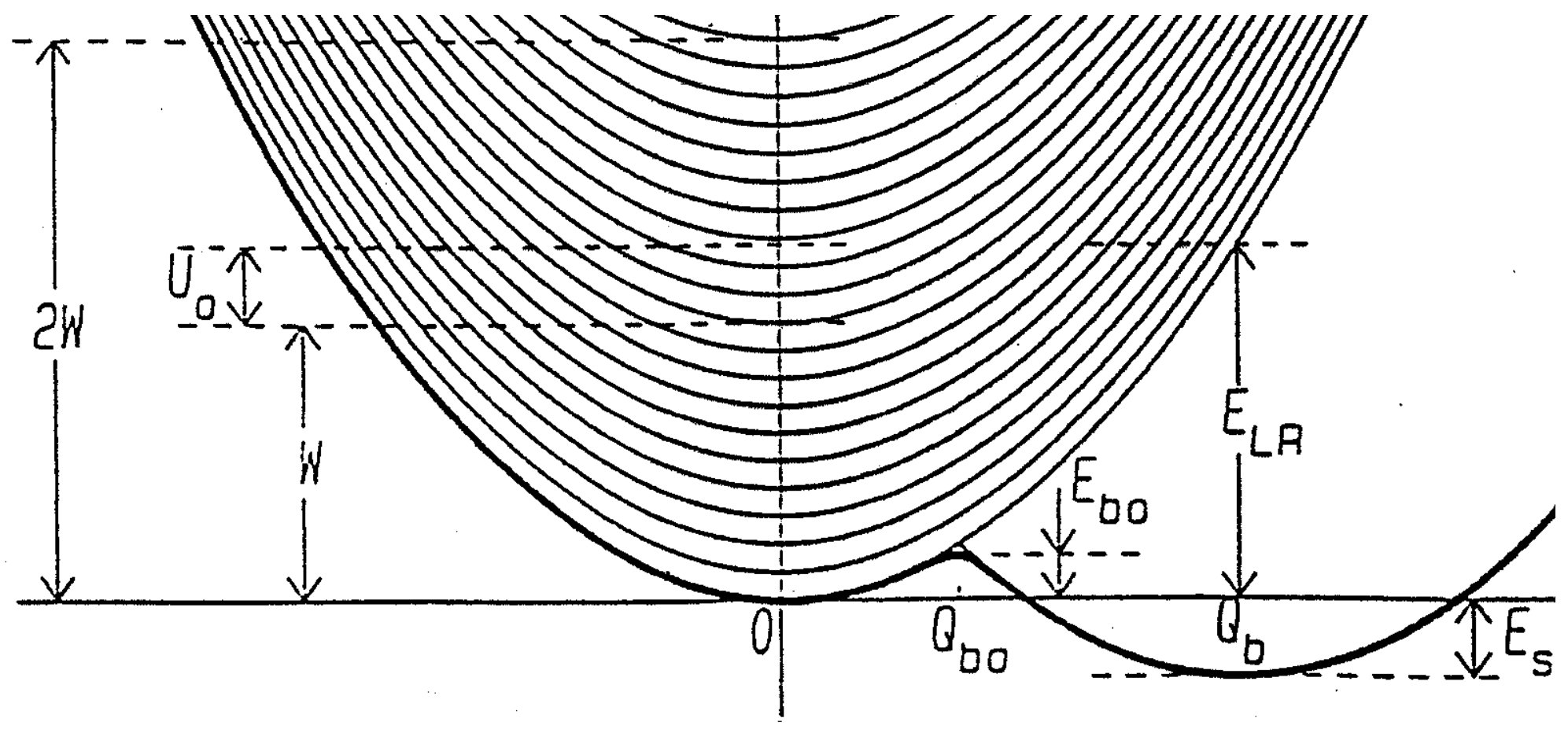

ig. 1 Adiabatic potentials of the sea of free holes (left) and of the bound hole

(c)

Figure 2:

(a)- coupling of the I&N and F centre semi-continuum potentials to a radial phonon mode (schematic) and (b)- the presumed atomic configurations for the insulating GaAs(110) surface. Following photo-generation, free holes are captured at active GaAs bonds, while free electrons are trapped at dangling bonds associated with alkaline adatoms at doped surfaces or with Ga host atoms at clean surfaces: This leads to phase segregation of the anti-pode photolytic products. Subsequent capture of a bi-hole at a GaAs surface bond brings about the virtual demolition of that bond [2], while the electronic counterpart undergoes surface bi-polaron formation [15]. Hole induced bond demolition creates a virtual ionic and neutral atom atmosphere (as in Kirilian's effect) out of an illuminated surface [16]. The electrostatic coupling at I&N and F center shows Coulomb's tail repulsive for I&N and attractive for F center. (c)- the phonon coupled potential gives rise to a sea of free hole vibronic states coupled electronically to a bound self trapped state.

## 2. Analytic background

Analytically, Itoh-Nakayama's potential comprises a flat bottom-well part followed by a screened Coulomb potential (*alias* Debye Hückel, *alias* Yukawa potential). For a two-component plasma, the Poisson-Boltzmann theory [4] of low screening potentials ($e\Phi \ll k_B T$) yields straightforwardly

$V(r) = -V_0 \ (r < r_0)$

$= +(e^2/\varepsilon r) \exp(-\kappa_D r) \ (r \geq r_0)$ (1)

where $\kappa_D(r) = r_D^{-1} = (8\pi N e^2/\varepsilon k_B T)^{1/2}$ is the reciprocal Debye screening radius, ε is an appropriate dielectric constant. Note that the Coulomb tail is repulsive, as in Figure 1.

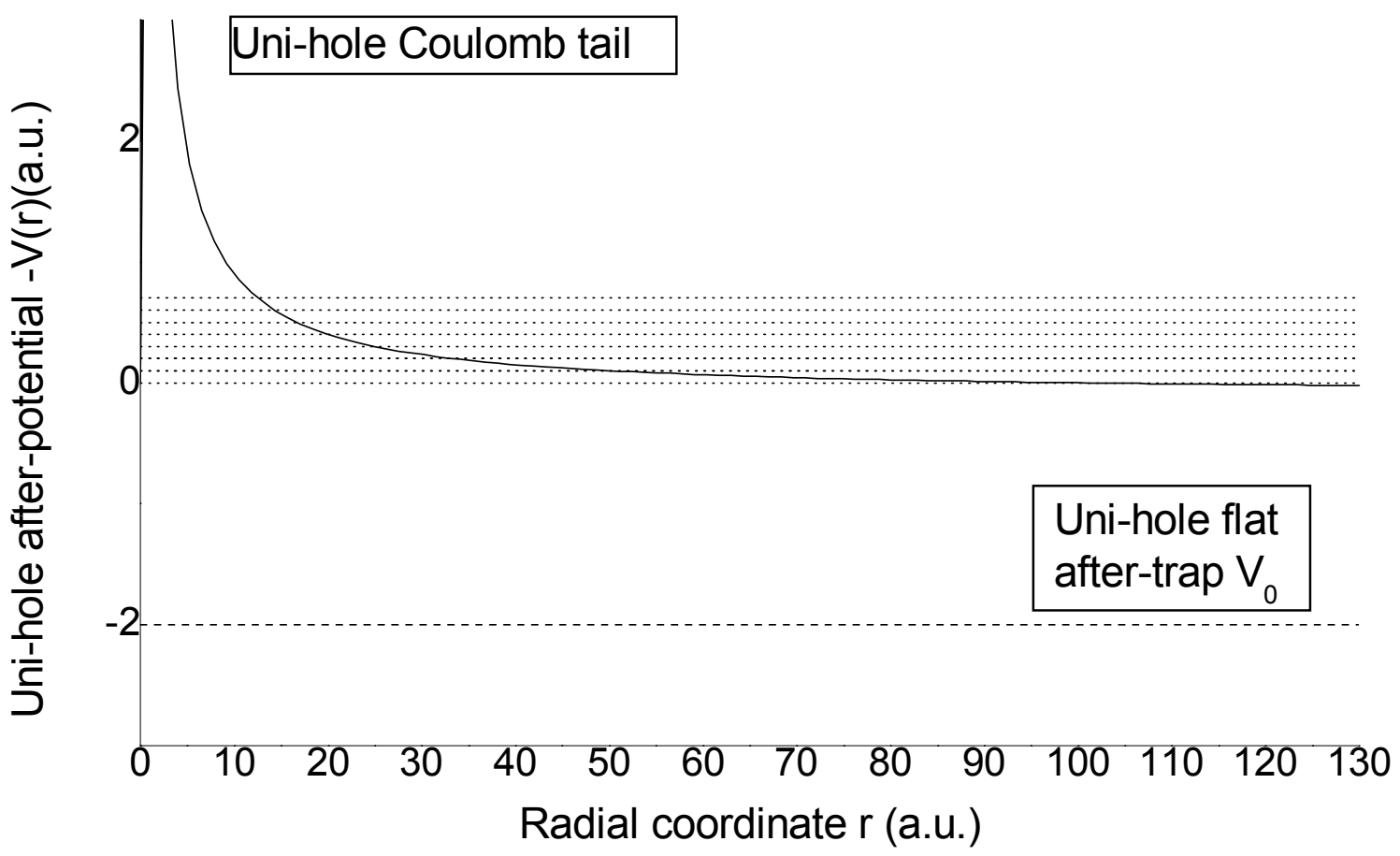

Figure 3:

The active width of the barrier is determined by the ultimate height of the energy level traversed along by the approaching band after-hole. The barrier width is thus controlled by an energy level of the Boltzmann tail statistics. In an alternative model labeled Fermi-Dirac, the ultimate height is determined by Pauli's exclusion principle at low temperatures. In the drawing above, we exemplify a system of 4 band states each contributing two band holes at most. (See references [2] and [14], respectively.)

Introducing the probability for traversing the barrier W(r) we define an average phonon coupled traversal rate [5]:

$K(E) = \int Z_0(E_n)^{-1} W(E_n) \exp(-E_n / k_B T) \, dE_n$ (2)

where $Z_0(E_n)$ is the partition function pertinent to the assumed statistics, while the integration is over the quantized energy levels generated by the screened potential (1):

$- (h^2/2M) \{[(d^2/dr^2) + [l(l+1)/r^2](d/dr)]\}\psi + V(r)\psi = E_n\psi$ (3)

The screened potential is quite significant at $r \sim r_D$, amounting to $V_D = V(r_D) = (e^2\kappa / 2.718\varepsilon)$ , which controls the I&N process for $r_0 << r_D$.

## 3. Probabilities for transition through barrier

John Bardeen [6] and then Stefan Christov [7] have elaborated an ingenious method for calculating the transition probabilities in barrier controlled processes such as in I&N's case. Under these conditions, the tunneling probability reads

$$W(E_n) = j_{\text{transmitted}} / j_{\text{incident}} \quad (4)$$

$$j(q) = ½\, i \sqrt{(h\nu/M)}\, [\chi d\chi^*/dq - \chi^* d\chi/dq] \quad (5)$$

Undoubtedly, all the underlying quantities can be found by solving Schrödinger's equation with a radial potential. We present the final results though the assumptions made are given below.

We have assumed that the motion along the configuration coordinate q is barrier controlled. In particular, the configuration transition probability along the radial coordinate based on currents across the barrier will be [6]:

$$W_{if\,conf}(E_n) = 4\pi^2\, |V_{fi}|^2\, \sigma_i(E_n)\sigma_f(E_n) \quad (6)$$

where the matrix element $V_{fi}$ is to be calculated using initial and final state wave functions $\phi_i$ and $\phi_f$, respectively, where $q_c$ is the crossover coorinate, as:

$$V_{fi} = (-h^2/2M)\, [\phi_f^* (d\phi_i/dq) - \phi_i (d\phi_f/dq)^*]|_{q=qc} \quad (7)$$

($h$ = h /2π). Here $\sigma_i$ and $\sigma_f$ are the corresponding density-of-states (DOS) of the initial and final states. For harmonic oscillators $\sigma_i(E_n) = \sigma_f(E_n) = (h\nu)^{-1}$. Inserting into (1) and performing the mathematics in (2) we obtain the relevant formulas for the text [7].

## 4. Solutions to the screening equations

For the numerical work we considered: (i) spherical Bessel functions at $r \le r_0$ inside the $V_0$ well [8], (ii) Coulomb functions (repulsive or attractive hydrogen-like ones) for $r_0 \le r \le r_D$ [9], and (iii) screened repulsive Coulomb functions for $r_D \le r < \infty$ [10]. The first one is well known, as is the second one, but the third one has only been tabulated numerically. The calculated rate for traversing the bi-hole sensed barrier is shown below in Figure 4 at the parameters indicated therein. The starting point for analyzing the screening problem is the complete Poisson-Boltzmann equation [11]. It displays both periodic and non-periodic solutions for the screened potential. The Debye equation is a reduced form derived for low potentials such that $|\Phi| \ll k_BT$. The following symmetries seem relevant to the present problem:

### 4.1. Spherical symmetry

Incorporating Laplace's operator $\Delta_{r,\theta,\varphi}$ in equation (2) and integrating over the angles, we obtain the Poisson-Boltzmann equation appropriate to spherical symmetry

$$(1/r^2)\,(d/dr)\,(r^2 d\Phi/dr) = \sinh\Phi \qquad (8)$$

or, equivalently,

$$d^2\Phi/dr^2 + (2/r)\, d\Phi/dr - \sinh\Phi = 0 \qquad (9)$$

The respective spherical symmetry Schrödinger equation reads (on applying the substitution $\Psi(\mathbf{r}) = [\psi(r)/r]\, Y_l^m(\theta,\varphi)$ as above):

$$(-h^2/2M)[d^2/dr^2 - l(l+1)/r^2]\psi - \Phi\psi = E\psi \qquad (10)$$

The electrostatic screening problem is solved through combining equations (10) and (9). We verify solving equation (9) at $\Phi \ll 1$ by means of $\Phi(r) = (e^2/r)\exp(-\kappa r)$ (Debye's solution) and indeed we find $d^2\Phi/dr^2 + (2/r)d\Phi/dr - \Phi = 0$ for $\Phi(r) = (e^2/r)\exp(-\kappa r)$.

For the complete potential $\sinh\Phi = \frac{1}{2}[\exp\Phi - \exp{-\Phi}] \neq \Phi$ and equation (9) holds good. (However, for $\sinh\Phi \sim \Phi$, the solution is as above.) By substitution $u = r\Phi$ we get $u'' + u = 0$ which is solved in $u(r) = A\exp(-\kappa r)$. (See Ref. [6] eqn. 2.101).

## 4.2. Circular cylindrical coordinates

In circular cylindrical coordinates, the Laplace operator reads:

$$\Delta_{r,\varphi,z}\,\Phi = d^2\Phi/dr^2 + (1/r)\,(d\Phi/dr) \qquad (11)$$

and the PB equation turns in:

$$d^2\Phi/dr^2 + (1/r)\,(d\Phi/dr) - \sinh\Phi = 0 \qquad (12)$$

Its linearized version is solved in cylindrical Bessel functions: $\Phi(r) \propto J_0(2\sqrt{[\lambda r]})$, whereas the eigenvalues obtain from $J_0(2\sqrt{\lambda}) = 0$. (See Ref. [6] eqn. 2.95). The solutions to various symmetry forms of the screening equations (P-B or D-H) are summarized in Table I. Note that there are no periodic D-H solutions, a genuine feature of the complete screening equation.

Table I
Solutions to the screening problem (Ref.[11])

| Symmetry | DH solution | Periodic PB solution | Aperiodic solution |
|---|---|---|---|
| spherical | $(1/r)\exp(-\kappa r)$ | $2\ln[\cot(\frac{1}{2}\kappa r)]$ (asymptotic) | $2\ln[\coth(\frac{1}{2}\kappa r)]$ |
| cylindrical | $J_0(2\sqrt{[\lambda r]})$ | $2\ln[\mathrm{cotan}(½\kappa r)]$ (asymptotic) | $\ln\{[(\kappa r)^2\sin^2(½\sqrt{H}\times \ln(r/r_0)]/H\}$ |
| planar | $\propto \exp(-\kappa x)$ | $2\ln[\mathrm{cotan}(½\kappa x)]$ (exact) | $2\ln[\mathrm{cotanh}(½\kappa x)]$ (exact) |

Table II
Oscillatory spherical well (See Ref.[13])

| 3D oscillatory spherical well | Eigenvalues | Eigenstates | Radial wavefunction |
|---|---|---|---|
| $V(r) = ½\, M\omega^2 r^2$ | $E_{nl} = h\omega(2n + l + 3/2)$ | $\psi_{nlm}(\xi) = (1/\xi)R_{nl}(\xi)Ylm(\theta,\varphi)$ $\xi = r/(h\omega/M\omega^2)$ | $R_{nl}(\xi) = N_{nl}\exp(-½\xi^2) \times\xi^{l+1}\mathbf{F}(-n,l+3/2,\xi^2)$ |

Here **F** is the degenerate hyper-geometrical function. Further discussion of the screened Coulomb wave functions can be found in Ref. [12].

### 4.3. Oscillatory spherical well

Deducing the electronic eigenvalues for a potential even as simple as the one of a spherical well is met with considerable mathematical difficulties [13]. The oscillatory well constitutes but another example of spherical symmetry in which the particle potential is dependent on the quadrate of the radial coordinate r rather than being independent of that coordinate, as in the previous example:

$$(-h^2/2M)[d^2/dr^2 - l(l+1)/r^2]\psi + ½M\omega^2r^2\psi = E\psi \qquad (13)$$

The eigenvalues of (13) remind of a three-dimensional harmonic oscillator [13]:

$$E_n = h\omega\,(n_x + n_y + n_z + 3/2) \qquad (14)$$

Whereas it is formidable deriving the eigenvalues of the flat spherical well, it is easy to deduce the ones of the 3D oscillator.

The oscillatory well is not associated with any obvious potential energy barrier. Nevertheless, this well may be considered remindful of the outer envelope of a double well potential composed of two steeper (twice as steep) configurational wells with a barrier in-between. The contents of Table II may then be regarded as the raw material for constructing an interwell barrier. The double well analogy may be found useful for solving for the interwell rates in corresponding situations.

### 4.4. Semi-continuous approach to F center potential

### (attractive Coulomb tail)

The F center (electron trapped at anion vacancy) plays a leading role amongst defects at trapped electron centers in solids which has taught solid state physicists how to tackle defect physics. Solving for the F center problem is not at all trivial for it involves taking into account polarization both electronic and ionic, as well as a wealth of vibronic effects to mention a few. The exact Schrödinger equation is therefore not easy to simplify and solve even by approximate methods. Nevertheses, in some instances replacing the realistic F center potential by a model potential that can be dealt with through approximations may prove fruitful leading to informative results. Such is the semi-continuous model which envisions the F center potential as a flat constant at short radial distances followed by an attractive Coulomb tai at longer ones (see the respective graph in Figure 1), e.g.

$V(r) = -V_0 \ (r < r_0)$

$= -(e^2/\varepsilon r) \exp(-\kappa_D r) \ (r \geq r_0)$

The semi-continuous F center has helped obtain fast and easy solutions in cases where there is no easy way through (see our earlier monograph [18] on F' centers in alkali halides). In a broader sense, it poses another example for a spherically symmetric trapped hole potential complementing the trapped hole potentials considered so far.

## 5. Numerical calculations

### 5.1. Flat spherical well

#### 5.1.1. Transitions between localized levels

Regarding the transitions from a spherical well level to a neighboring localized level we apply directly the arguments leading to Section 4. For a flat spherical well we take the aperiodic solution from Table I to set

$\phi_i(r) = 2\ln[\mathrm{cotanh}[\tfrac{1}{2}\kappa(r-\delta)]],$

$d\phi_i / dr = 2/[\mathrm{cotanh}[\tfrac{1}{2}\kappa(r-\delta)]][1/\sinh^2[\tfrac{1}{2}\kappa(r-\delta)]] = 2\kappa/\cosh^2[\tfrac{1}{2}\kappa(r-\delta)]$

$\phi_f(r) = 2\ln[\mathrm{cotanh}[\tfrac{1}{2}\kappa(r+\delta)]],$

$d\phi_f / dr = - 2\kappa/\cosh^2[\tfrac{1}{2}\kappa(r+\delta)]$

at $\delta \to 0$. Equations (6) and (7) then yield at the barrier peak $r_b$:

$$V_{if}(r_b) = 2 \times 2\kappa/\cosh^2[\tfrac{1}{2}\kappa(r_b-\delta)] \times 2\ln[\mathrm{cotanh}[\tfrac{1}{2}\kappa(r_b+\delta)] = 8\ \kappa/\cosh^2(\tfrac{1}{2}\kappa r_b) \qquad (15)$$

$$W_{if}(r) = 4\pi^2\ (h^2/2M)\ |\ 8\ \kappa/\cosh^2(\tfrac{1}{2}\kappa r_b)|^2\ (1/\sigma_i)(1/\sigma_f)\ |_{\ r = rb} \qquad (16)$$

An alternative approach to transitions between localized levels can be found in Ref. [7].

### 5.1.2. Transitions from band to localized levels (self trapping rates)

The band constitutes a sea of quantized energy which may be visualized along the energy axis as an array of levels equidistant at $dE_g(E_n)$. In as much as the energy gaps $dE_g(E_n)$ between neighboring levels are infinitesimal, new methods have been elaborated in that we first define a differential rate proportional to $dE_g(E_n)$ and then integrate to obtain the integrated rate. A detailed analysis having already been published elsewhere [2], we will only reproduce the essentials below.

The differential band-to-local-level rate reads

$$d\Re = (4\pi/h^2\omega)\sqrt{(\pi k_B T/E_{LR})}\ |H_{b\mathbf{k}}(Q)|^2\ \sinh(h\omega/2k_B T)\ \exp(-E_{b\mathbf{k}}/k_B T) \qquad (17)$$

where $E_{LR}$ is the electron-phonon coupling energy, $h\omega$ is the coupled phonon quantum, $H_{b\mathbf{k}}(Q)$ is the off-diagonal matrix element of the interaction Hamiltonian. Averaging over the band energies we get:

$$\Re(\varepsilon) = (\pi/h^2\omega)\sqrt{(\pi k_B T/E_{LR})}\sinh(h\omega/2k_B T)\ {}_{\varepsilon}\!\int^{\infty}(\varepsilon-E)^2\exp[-(\varepsilon-E)^2/4E_{LR}k_B T)]\rho(E)dE \ (18)$$

where $U_0$ is the absolute well depth, $\rho(E)$ is the DOS of free-hole states, W is the hole conduction-band half-width. We get after integrating [Gradstein & Ryzhik (p. 351)]:

$$\Re(\varepsilon) = (\pi/h^2\omega)\ \pi(k_B T)^2\ \sinh(h\omega/2k_B T)(N/W)\ -$$

$$(\pi/h^2\omega)\sqrt{(\pi k_B T/E_{LR})}\ \sinh(h\omega/2k_B T)\ {}_0\!\int^{\varepsilon}(\varepsilon-E)^2\exp[-(\varepsilon-E)^2/4E_{LR}k_B T)](N/W)dE \quad (19)$$

where here and above $\varepsilon = (W+U_0)$. The first term in eqn. (19) $\propto (k_B T)^2$ yields $\Re(0)$:

$$\Re(0) - \Re(\varepsilon) = (\pi/h^2\omega)\sqrt{(\pi k_B T/E_{LR})}\ \sinh(h\omega/2k_B T)\ \times$$

$${}_0\!\int^{\varepsilon}(\varepsilon-E)^2\exp[-(\varepsilon-E)^2/4E_{LR}k_B T)](N/W)dE \qquad (20)$$

At high enough temperatures $2k_B T\ /\ h\omega \gg 1$, equation (20) transforms into an explicit temperature dependence of the form

$$\Re(\varepsilon) = (\pi/h^2\omega)\pi(k_B T)^2\sinh(h\omega/2k_B T)(N/W) - (\pi/h^2\omega)\sqrt{(\pi k_B T/E_{LR})}\sinh(h\omega/2k_B T)I(\varepsilon)(N/W),$$

where

$I(\varepsilon) = {}_0\int^{\varepsilon} (\varepsilon\text{-}E)^2 \exp[-(\varepsilon\text{-}E)^2/4E_{LR}k_BT)]dE$

$= {}_{E=0}\int^{E=\infty} (\varepsilon\text{-}E)^2/ (4E_{LR}k_BT) \exp[-(\varepsilon\text{-}E)^2/4E_{LR}k_BT)]\ d\surd[(-E/4E_{LR}k_BT)](4E_{LR}k_BT)^{3/2},$

as follows:

$\Re(\varepsilon) = \omega(\pi^2/2)(k_BT/h\omega)(N/W) - \omega(1/2h\omega)(\pi^{3/2})\ I(\varepsilon)\ /\surd(E_{LR}\ k_BT)(N/W)$

$= \omega\ (\pi^{3/2}/2)(1/h\omega)[\surd\pi\ (k_BT) - I(\varepsilon)\ /\surd(E_{LR}\ k_BT)](N/W)$

with an activation energy of ~ a phonon quantum. Here the integral is expressed in terms of transcendent functions as follows:

$I(\varepsilon) = (4E_{LR}k_BT)^{3/2}\ {}_0\int^{\varepsilon / \surd(4ELRkBT)} \lambda^2 \exp(-\lambda^2)d\lambda$

$= (4E_{LR}k_BT)^{3/2}\{\varepsilon /\surd(4E_{LR}k_BT)\ [1 - \exp(-\ \varepsilon /\surd(4E_{LR}k_BT))] + \mathrm{erf}\ (\varepsilon /\surd(4E_{LR}k_BT\ ))\}$ (21)

## 5.2. Spherical oscillator

From equations (13) and (14) making use of the oscillatory eigenfunctions

$\phi_i(r) = A_i \exp(-a(r\text{-}\delta)^2)$

$d\phi_i /dr = -A_i\, 2a(r\text{-}\delta)\exp(-a(r\text{-}\delta)^2 )$

$\phi_f (r) = A_f \exp(-a(r+\delta)^2 )$

$d\phi_f /dr = -A_f\, 2a(r+\delta)\ \exp(-a(r+\delta)^2 )$

at $\delta \to 0$; equations (6) and (7) at the barrier peak $r_b$ coordinate yielding for a 3D oscillator in ground state:

$V_{if} (r) =(-h^2/2M)A_iA_f\{-2a(r+\delta)\exp(-a(r+\delta)^2 )\exp(-a(r\text{-}\delta)^2 )-2a(r\text{-}\delta)\exp(-a(r\text{-}\delta)^2 )\exp(a(r+\delta)^2\}$

$= (-h^2/2M)A_iA_f \exp(-a(r+\delta)^2 )\exp(-a(r\text{-}\delta)^2 )[\{-2a(r+\delta) - 2a(r\text{-}\delta)]$

Now that the barrier-current potential reduces to

$V_{if}(r) = (-h^2/M)A_iA_f \exp(-a(r+\delta)^2 )\exp(-a(r\text{-}\delta)^2 )\ (-2a)\ r$ (22)

and, consequently, the barrier-transition probability is found to be

$W_{if}(r) = 16a^2\pi^2(h^2/2M )^2 \mid A_iA_f \exp(-a(r_p+\delta)^2) \exp(-a(r_p\text{-}\delta)^2) \mid^2 r_p\,(1/\omega_i)(1/\omega_f) \mid_{r = rb}$ (23)

at $\delta \rightarrow 0$. Note that simple oscillator models usually assume $\omega_i = \omega_f$ which is not necessarily the present case. The oscillating spherical well story is tabulated in Table II.

Considerations similar to those leading to equation (19) can be carried out resulting in an integrated rate for the oscillating spherical well akin to the flat well. Combined they both tell of an universal property characteristic of the spherical symmetry.

## 5.3. Suggested parameters and graphics for future studies

Two sets of calculations were also considered (not shown): one related to a spherical well (a) and the other to a spherical oscillator (b). Comparable parameters were used, most important among them being the Debye screening radius $r_D = [8\pi ne^2/\kappa k_B T]^{-1/2}$ calculated to $r_D = 59$ Å at $n = 10^{17}$ /cc (free electron concentration), $\varepsilon = 5$ (dielectric constant) and T = 300 K. To be on the safe side, $r_0$ was assumed to be one tenth of the Debye radius estimate (6 Å). Presently we have no policy on the angular frequencies to be incorporated; for this reason we set $\omega_i \sim \omega_f \sim 10^{13}$ $s^{-1}$. We also set $A \equiv N_n = [\sqrt{\pi} n! 2^n]^{-1/2}$ and $a = \sqrt{½}\sqrt{(M\omega^2/h\omega)}$ from textbooks on quantum mechanics [8].

## 6. Concluding remarks

Theories outlined above apply to barrier controlled processes in solid and, possibly, nuclear matter as well. Such is the hole transfer across I&N's potential where a semi-barrier forms near the active bond boundary at $r = r_0$ which is a semi-barrier in the sense that only its outside form is described by a smooth radial potential function while its inner part is abrupt. Unlike it, barrier forms appearing in diffusion, nucleation, etc. are smooth on both sides of a boundary. Full forms often occur in nature, while semi-forms occur in artificial material structures such as ones in junctions, etc.

Unlike the above, the oscillatory well is not barrier controlled, its extremum appearing at $r = r_0$ is one of minimum, as typical for an oscillator. Nevertheless, we included the spherical well oscillator in the present discussion for the simple reason that it complements the barrier case by appearing as its antipode. There is another simple reason in this same context, as the oscillatory well is less popular among solid state albeit not nuclear physicists.

Undoubtedly, the oscillatory well may be extended so as to comprise the nonlinear species, as described by Mathieu's transcendent functions [9]. As Mathieu's extension relates to the linear oscillator, so would our conceived extension relate to its linear spherical analogue.

One way or the other, the conclusions that can be drawn on ground of our statistical approach described presently as regards the sputtering rate agree generally with the ones based on the conventional approach [2]. This is not surprising since common elements have been used in both cases.

The general trend of the statistical I&N rate shown in Figure 4 is typical for thermally-activated rate processes. In it the barrier stipulates an activation energy for the temperature dependence. However, there is another dependence on the excitation light intensity I by laser power (fluence) $\Im$ which points even more directly to the role of the screening process. This dependence is reproduced in Figure 5 where a bunch of experimental points is shown along with a deduced dependence of the statistical rate on the density of band electrons n(I). In fact n(I) enters into the basic equations $\propto \exp(-\kappa_D r_0)$ through the reciprocal screening radius $\kappa_D = (8\pi n(I)e^2/\varepsilon k_B T)^{1/2}$. The graph points to the occurrence of a threshold fluence $\Im_C$ which controls sputtering in that the commanding barrier is too high for $\Im < \Im_C$ and that it drops to open up the process for $\Im \geq \Im_C$. We remind that the control over threshold is exerted by the excitation light intensity via n(I).

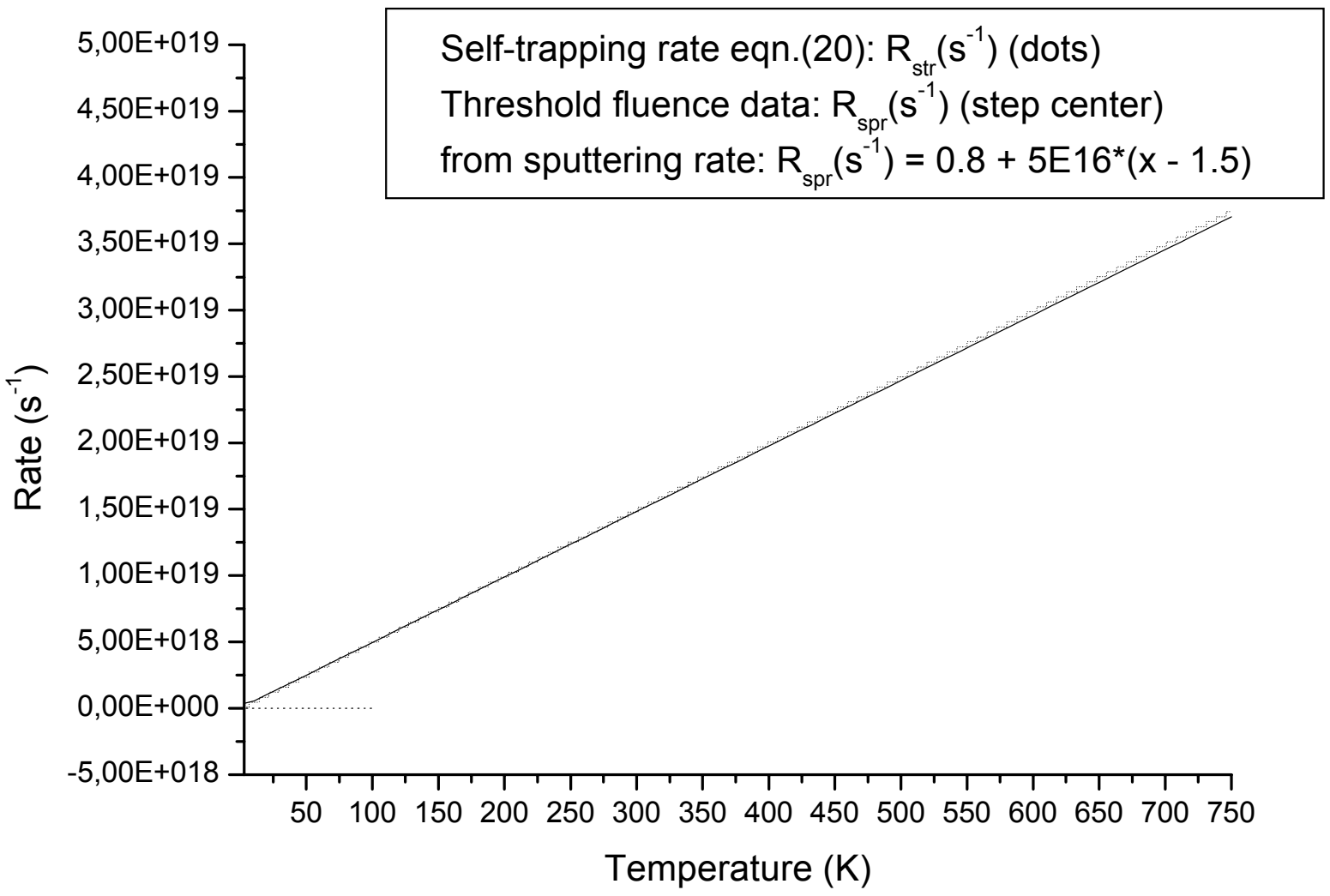

Figure 4:

Calculated temperature dependence of the rate for traversing the uni-hole after-barrier by a band hole. Also shown as a self-trapping rate from equation (20), it is typical for a band-to-localized-level transfer. The employed parameters are: hole (half)-bandwidth W = 2.5 eV, hole spherical well depth $U_0 = - V_0 =$ 0.5 eV, coupled phonon quantum $\hbar\omega$ = 25 meV, lattice relaxation energy $E_{LR}$ = 0.25 eV. A good accord is on display between the theoretical self-trapping rate $R_{str}$ (dots) and the experimental threshold fluence rate $R_{sr}$ (step center).

The subsequent capture of a dihole at a GaAs bond brings about the virtual demolition of that bond [2], while the electronic counterpart undergoes surface bipolaron formation [15] through di-electron capture by dangling bonds [15]. The hole produced bond demolition creates virtual ionic or neutral atom atmosphere out of an illuminated surface [16]. In the former case the surface becomes electrically charged to compensate for the electrostatic cloud outside. The whole bond demolition undergoing body will become electrified.

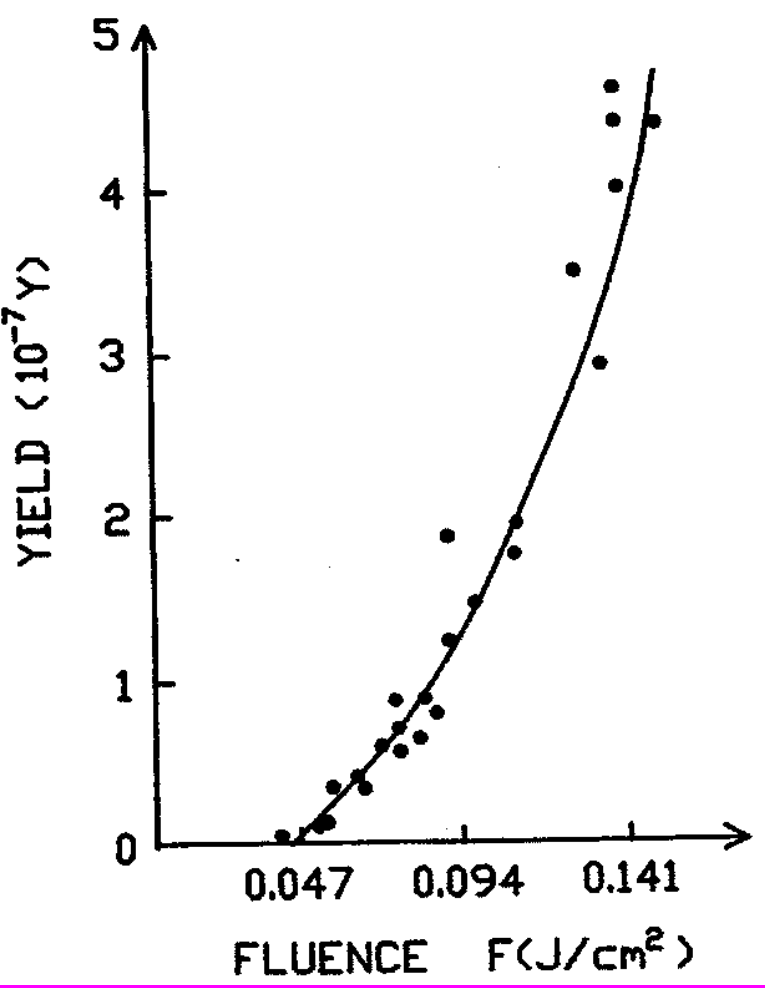

(a)

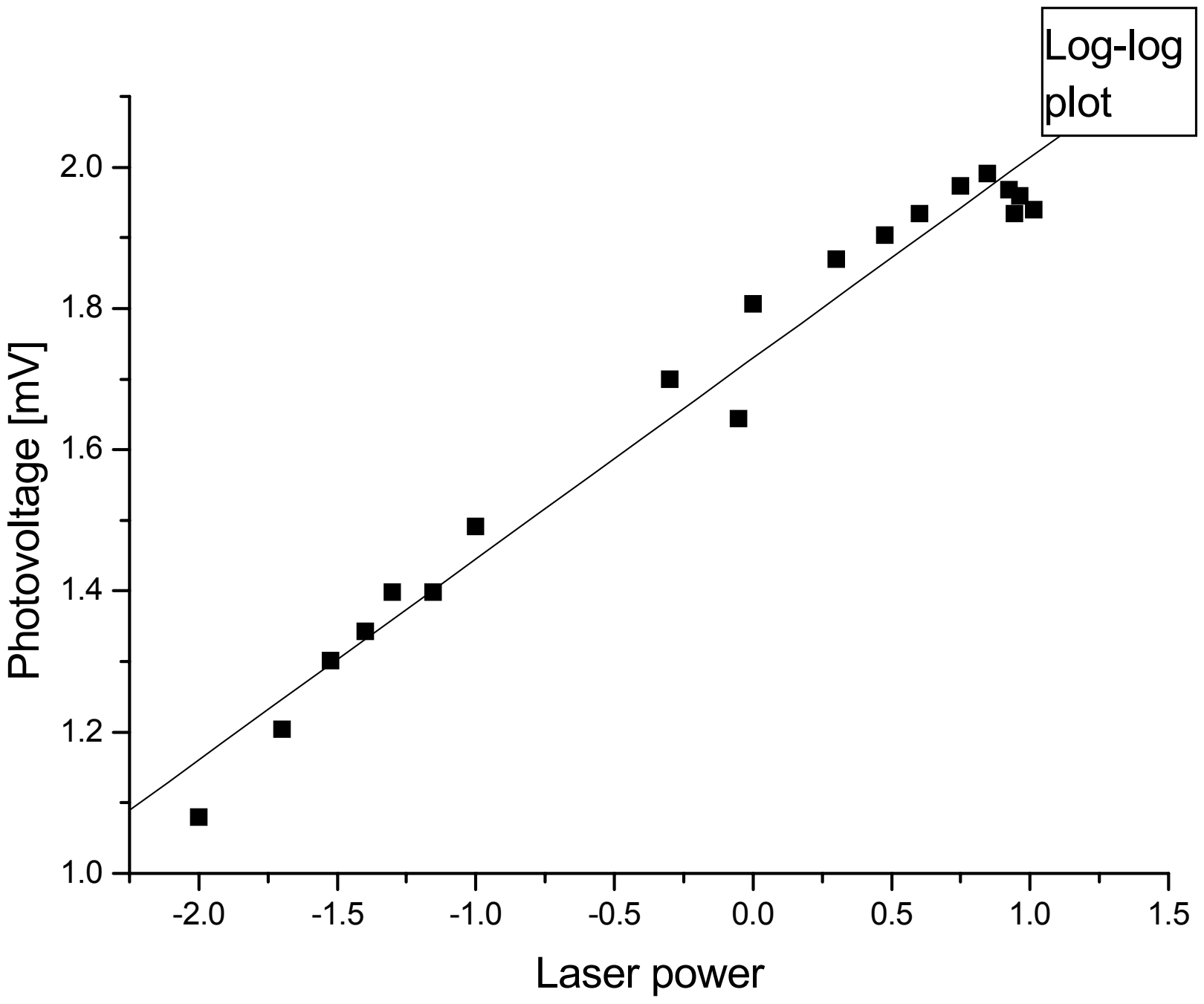

(b)

Figure 5:

The dependence of self-trapping rate $\Re$ on laser power (fluence) $\Im$, points directly to the role of screening. We note the following: (a)- kinetics of sputtering Ga neutrals from (110)GaP surface at room temperature (after Ref.[2]). The fluence dependence in a laser sputtering experiment depends on the form of the screening potential $\propto (1/r)\exp[-\kappa_D(I)r]$ where $\kappa_D(\Im) = \sqrt{[8\pi n(\Im)e^2/\varepsilon k_B T]}$. Apparently, $n(\Im)$ enters into the basic equations through the reciprocal screening radius $\kappa_D(\Im)$ which in its turn makes the electron-phonon coupling constant $\beta(r)$ and the coupling energy $E_{LR}$ doping dependent [2]. The graph points to a threshold fluence $\Im_C$ to control sputtering, in that the barrier is too high freezing in the sputtering for $\Im < \Im_C$ while it drops low enough to switch in the process at $\Im \geq \Im_C$. The obtained exponential form of $\Im$ dependence suggests $n(\Im) \propto \Im$ within the sputtering range. (b)- laser power dependence of the photo-electrification rate at a (111)GaAs surface under the conditions of a synchronous-detection experiment (after Ref. [16]). Its roots in the logarithmic plot above are more or less evident.

In summary, closely related to the sputtering of ionic particles is the photo-electrification of insulating crystals [16]. The process creates a charged ion atmosphere outside and all-around the illuminated body while the surface becomes charged to compensate for the atmospheric charge. Time resolved and temperature experiments have demonstrated the feasibility of our interpretation.

The obtained exponential-like form of intensity dependence suggests $n(I) \sim I$ within the sputtering range in $A_{III}B_V$. A related intensity dependence is obtained for the photo-electrification rate in other materials [16]under different conditions.

Further studies into the sputtering problem including some new analytic techniques and results complementing reference [2] can also be found in the literature [17].

Acknowledgements

I am grateful to Professor Noriaki Itoh (Nagoya) for an invitation to his laboratory and data necessary for constructing the theory in Ref. [2]. Talks with Y. Nakai and K. Hattori are also appreciated greatly. Later Professor Jai Singh (Darwin) provided the working atmosphere for completing our excursion across laser sputtering.

References

[1] N. Itoh, and T. Nakayama, Phys. Letters **92A**, 471-475:1982.
[2] M. Georgiev and J. Singh, Appl. Phys. A - Solids & Surfaces **55** (2) 170-175 (1992).
[3] J. Singh, N. Itoh and V.V. Truong, Appl. Phys. A **49** 631 (1989):
[4] M. Georgiev, arXiv:0901.0717.
[5] M. Georgiev, arXiv:0903.0856.
[6] J. Bardeen, Phys. Rev. Lett. **6** (2) 57 (1961):
[7] S.G. Christov, *Collision Theory and Statistical Theory of Chemical Reactions Lecture Notes of Physics* #18 (Springer, Berlin, 1980).
[8] L.I. Schiff, *Quantum Mechanics* (McGraw Hill, NewYork, 1968).
[9] M. Abramowitz in: M. Abramowitz and I.A. Stegun, eds.: *Handbook of mathematical functions with formulas, graphs and mathematical tables* (NBS, Applied Math series, 1964), p. 354: "*Coulomb wave functions*." (Russian translation: (Moskva, Nauka, 1979).
[10] N. Martinov, D. Ouroushev and E. Chelebiev, J. Phys. A: Math. Gen. **19**, 1327 (1986).
[11] A.J.M. Garrett and L. Poladian, *Annals of Phys. (NY)* **188** (2) 386-435 (1988).
[12] V.L. Bonch-Bruevich and V.B. Glasko, *Optika i Spektroskopiya* **14** (4) 495 (1963).
[13] A.S. Davydov, *Quantum Mechanics* (GIFML, Moscow, 1963).
[14] Hitoshi Sumi, Surface Sci. **248**, 382-400 (1991).
[15] O. Pankratov and M. Scheffler, Phys. Rev. Letters **71** (17) 2797 (1993).
[16] O. Ivanov, R. Dyulgerova and M. Georgiev, cond-mat/0508457: cond-mat/0508460; 0706.3877; 0805.0013.
[17] M. Georgiev, L. Mihailov, J. Singh in: *Electronic, Optoelectronic and Magnetic Thin Films*, J.M. Marshall, N. Kirov and A. Vavrek, eds. (Research Studies Press Ltd. and John Wiley & Sons, Inc., Taunton and New York, 1995), p. 507.
[18] Mladen Georgiev, *F' Centers in Alkali Halides*. Lecture Notes in Physics #80 (Springer, Berlin-New York, 1989).